\def\proof{\noindent{\bf Proof: }}

\def\x{{\bf{x}}}

\newcommand{\argmin}{\operatornamewithlimits{argmin}}
\newcommand{\Tr}{\operatornamewithlimits{Tr}}

\def\qed{\hfil\hfil{\vrule height7pt width6pt depth0pt}\hfil}

\documentclass[12pt]{article}
\newtheorem{theorem}{Theorem}[section]

\newtheorem{lemma}[theorem]{Lemma}

\newtheorem{remark}[theorem]{Remark}

\usepackage{amssymb}
\usepackage{amsmath}
\usepackage[pdftex]{hyperref}
\usepackage{graphicx}
\usepackage{bm}
\usepackage{epstopdf}

\title{On the Consistency of Compressed Modes for Variational Problems}

\author{ 
Farzin Barekat\thanks{Mathematics Department, University of California at Los Angeles,
Los Angeles, CA 90095-1555 USA. fbarekat@math.ucla.edu.} ~
}

\date{\ }

\begin{document}
\maketitle

\begin{abstract}
This paper provides theoretical consistency results for compressed modes introduced in \cite{BIB:CMpaper1}. We prove that as $L^1$ regularization term in certain non-convex variational optimization problems vanishes, the solutions of the optimization problem and the corresponding eigenvalues converge to Wannier-like functions and the eigenvalues of the Hamiltonian, respectively. 
\end{abstract}

\section{Introduction}\label{introduction}
In \cite{BIB:CMpaper1,BIB:CMpaper2} the authors pioneered a new methodology of using sparsity techniques to obtain localized solutions to a class of problems in mathematical physics that can be recast as variational optimization problems.
The typical method used for finding orthogonal functions that span eigenspace of a Hamiltonian and are spatially localized is to choose a particular unitary transformation of the eigenfunctions of the Hamiltonian. In solid state physics and quantum mechanics these functions are known as Wannier functions \cite{BIB:Wannier}. There are many approaches to finding the ``right'' unitary transformation. The most widely used approach for calculating maximally localized Wannier functions (MLWFs) was introduced in \cite{BIB:MLWF}. There are two difficulties associated with this approach. First, the eigenfunctions of the Hamiltonian must be calculated. Second, one needs to determine a distance to manually cut off the resulting MLWFs.

In \cite{BIB:CMpaper1}, it was demonstrated that introducing $L^1$ regularization in the variational formulation of the Schr\" odinger equation of quantum mechanics and solving the new non-convex optimization problem, results in a set of localized functions called compressed modes (CMs). It was shown numerically that CMs have many desirable features, for example, the energy calculated using CMs approximates the ground state energy of the system. Moreover, there is no requirement to cut off the resulting CMs ``by hand''. In a more recent development, the ideas of \cite{BIB:CMpaper1} were used in \cite{BIB:CMpaper2} to generate a new set of spatially localized orthonormal functions, called compressed plane waves (CPWs), with multi resolution capabilities adapted for the Laplace operator.

The idea of using the $\ell_1$ norm as a constraint or penalty term to achieve sparsity has attracted a lot of attention in a variety of fields including compressed sensing \cite{BIB:compressedSensing1,BIB:compressedSensing2}, matrix completion \cite{BIB:rankMinimization}, phase retrieval \cite{BIB:phaseRetrieval}, etc. Recently, sparsity techniques began being used in physical science (see for example \cite{BIB:efficientMaterial}) and partial differential equations (see for example \cite{BIB:multiScale}). In all these examples sparsity means that in the representation of a corresponding vector or function in terms of a well-chosen set of modes (ie. a basis or dictionary), most coefficients are zero. However, \cite{BIB:CMpaper1,BIB:CMpaper2} for the first time advocates the use of $L^1$ norm regularization to achieve spatial sparsity (i.e. functions that are spatially localized).

In this paper we prove consistency results for compressed modes (CMs) introduced in \cite{BIB:CMpaper1}. In particular, we show that as $\mu\rightarrow \infty$ the approximate energy calculated using CMs converges from above to the actual energy of the system. This is done in section \ref{section:energies}. More importantly, in section \ref{section:eigenfunction}, we show that under some necessary assumptions on the spectrum of the Hamiltonian, as $\mu\rightarrow \infty$, CMs converge to a unitary transformation of the eigenfunctions (i.e. Wannier-like functions) in $L^2$ norm. Moreover, we verify a conjecture stated in \cite{BIB:CMpaper1}. 

Let $\hat{H}=-\frac{1}{2}\Delta+V(\x)$ denote Hamiltonian of a system with eigenfunctions $\phi_1,\phi_2,\ldots,$ and corresponding eigenvalues $\lambda_1\leq \lambda_2 \leq\ldots$. 
Observe that $\phi_1,\ldots,\phi_N$ are a solution to the optimization problem:
\begin{equation} \{\phi_1,\ldots,\phi_N\}=\argmin_{\tilde{\phi}_1,\ldots,\tilde{\phi}_N}\sum_{i=1}^{N}\langle\tilde{\phi_i},\hat{H}\tilde{\phi}_i\rangle \qquad \hbox{ s.t. } \qquad \langle\tilde{\phi}_j,\tilde{\phi}_k\rangle=\delta_{jk},
 \label{equation:propertyPhi}\end{equation}
where $\langle \phi_j,\phi_k\rangle=\int_{\Omega} \phi^*_j(\x)\phi_k(\x) d\x$. Throughout this paper we assume that domain $\Omega$ is a bounded subset of $\mathbb{R}^d$ with appropriate boundary conditions for the Laplacian.

In \cite{BIB:CMpaper1}, compressed modes $\{\psi_i\}_{i=1}^{N}$, corresponding to number $N$, are defined as the solution to $L^1$ regularized optimization problem 
\begin{equation} \{\psi_1,\ldots,\psi_N\}=\argmin_{\tilde{\psi}_1,\ldots,\tilde{\psi}_N}\sum_{i=1}^{N}\frac{1}{\mu}\|\tilde{\psi_i}\|_1+\langle\tilde{\psi_i},\hat{H}\tilde{\psi}_i\rangle \qquad \hbox{ s.t. } \qquad \langle\tilde{\psi}_j,\tilde{\psi}_k\rangle=\delta_{jk}, 
\label{equation:propertyPsi}\end{equation}
where $\|\psi_i\|_1=\int_{\Omega}|\psi_i|d\x$. As shown in \cite{BIB:CMpaper1,BIB:CMpaper2}, compressed modes have many desirable features. In particular, consider the $N\times N$ matrix $\langle \Psi_N^T,\hat{H}\Psi_N\rangle$ with the $(j,k)$-th entry defined by $\langle \psi_j,\hat{H}\psi_k\rangle$ and let $(\sigma_1,\ldots,\sigma_N)$ denote its eigenvalues in non-decreasing order. In \cite{BIB:CMpaper1,BIB:CMpaper2}, it was conjectured that as $\mu\rightarrow \infty$, $\sigma_i$'s converge to $\lambda_i$'s. In theorem \ref{theorem:mainEigenvalue} we verify this conjecture. We also show that as $\mu\rightarrow \infty$, CMs $\psi_i$'s converge to a unitary transformation of eigenfunctions $\phi_i$'s (i.e. Wannier-like functions) in the $L^2$ norm. Observe that orthonormality constraints in optimization problems \eqref{equation:propertyPhi} and \eqref{equation:propertyPsi}, renders them to be non-standard. In particular note that the space of feasible functions in \eqref{equation:propertyPhi} and \eqref{equation:propertyPsi} is not a convex set and many convex optimization techniques and analysis cannot be applied here.

Indeed, we show the results hold in a more general setting: Suppose $J:L^2\rightarrow \mathbb{R}^+$ is a nonnegative bounded operator on space of $L^2(\Omega)$ functions; that is, there exists a constant $C$ such that 
\[  0\leq J(g)\leq C\|g\|_2 \qquad \hbox{ for all }\qquad g\in L^2. \] 
 Let $\{f_i\}_{i=1}^{N}$ denote a set of solutions, corresponding to the number $N$, of the optimization problem 
\begin{equation} \{f_1,\ldots,f_N\}=\argmin_{\tilde{f_1},\ldots,\tilde{f_N}}\sum_{i=1}^{N}\frac{1}{\mu}J(\tilde{f_i})+\langle\tilde{f_i},\hat{H}\tilde{f}_i\rangle \qquad \hbox{ s.t. } \qquad \langle\tilde{f}_j,\tilde{f}_k\rangle=\delta_{jk}. 
\label{equation:propertyF}\end{equation}
Let $\langle F_N^T,\hat{H}F_N\rangle$ denote the $N\times N$ matrix whose $(j,k)$-th entry is $\langle f_j,\hat{H}f_k\rangle$ and let $(\nu_1,\ldots,\nu_N)$ denote its eigenvalues in non-decreasing order. Define the energy associated with this solutions by 
\begin{equation} 
E=\sum_{i=1}^{N}\langle  f_i,\hat{H}f_i\rangle=\Tr(\langle F_N^T,\hat{H}F_N\rangle)=\nu_1+\cdots+\nu_N. 
\label{equation:sumF}  \end{equation}

Recall that the domain $\Omega$ is bounded; in particular, operator $J(\cdot)=\|\cdot\|_1$ satisfies the above conditions. Therefore, results shown for solutions $f_1,\ldots,f_N$ and the corresponding $\nu_1,\ldots,\nu_N$, in particular hold for CMs $\psi_1,\ldots,\psi_N$ and corresponding $\sigma_1,\ldots,\sigma_N$.  Nevertheless, the case $J(\cdot)=\|\cdot\|_1$ is the most interesting application and the main motivation for considering this problem in the first place.

The remainder of this paper is as follows. In section \ref{section:energies}, we show that $E$ converges from above to the actual energy of the system as $\mu\rightarrow \infty$. Section \ref{section:eigenfunction} contains the main results of the paper; that is, as $\mu\rightarrow \infty$, $f_i$'s converge to a unitary transformation of $\phi_i$'s and eigenvalues $(\nu_1,\ldots,\nu_N)$ converge to $(\lambda_1,\ldots,\lambda_N)$. 
In section \ref{section:conclusion}, we make some concluding remarks. 

\section{Convergence of Energies}\label{section:energies}
In this section we show that $E$
converges to the ground state energy $E_0=\sum_{j=1}^N \lambda_j$ as $\mu\rightarrow \infty$. Although, the result of this section can be readily deduced from theorem \ref{theorem:mainEigenvalue}, we have included it here for its independent interest and simplicity of argument.

First observe that 
\begin{equation}  
E_0=\lambda_1+\cdots+\lambda_N=\sum_{i=1}^{N}\langle \phi_i,\hat{H}\phi_i\rangle \leq \sum_{i=1}^{N}\langle f_i,\hat{H}f_i\rangle=\nu_1+\cdots+\nu_N=E, 
\label{equation:sumLambdaSumSigma}\end{equation}
where we used property \eqref{equation:propertyPhi} for the inequality and equation \eqref{equation:sumF} for the last equality.

Next choose $\mu$ large enough such that 
\[ \frac{1}{\mu}\sum_{i=1}^{N}J(\phi_i) <\epsilon. \]
This is plausible due to boundedness of operator $J$ and the assumption that $\|\phi_i\|_2=1$ for $i=1\ldots,N$.

We have 
\begin{align}
 E=\nu_1+\cdots+\nu_N  =  \sum_{i=1}^{N}\langle f_i,\hat{H}f_i\rangle \leq  \frac{1}{\mu}\sum_{i=1}^{N}J(f_i)+ \sum_{i=1}^{N}\langle f_i,\hat{H}f_i\rangle  \notag \\
\leq  \frac{1}{\mu} \sum_{i=1}^{N}J(\phi_i)+ \sum_{i=1}^{N}\langle \phi_i,\hat{H}\phi_i\rangle < \epsilon+\lambda_1+\cdots+\lambda_N=\epsilon+E_0,  
\label{equation:sumSigmaSumLambda}\end{align}
where we used \eqref{equation:sumF} for the first equality, positivity of operator $J$ for the first inequality, and property \eqref{equation:propertyF} for the second inequality.

From equations \eqref{equation:sumLambdaSumSigma} and \eqref{equation:sumSigmaSumLambda} it follows that 
\[ E \downarrow E_0 \qquad \hbox{ as } \mu\rightarrow \infty.\]

\section{Consistency Results for $L^1$ Regularization}\label{section:eigenfunction}
This section contains the main results of the paper. In theorem \ref{theorem:main}, we show that as  $\mu\rightarrow \infty$, the solutions to the regularized optimization problem \eqref{equation:propertyF} converge to a unitary transformation of the eigenfunctions in $L^2$ norm. Furtheremore, in theorem \ref{theorem:mainEigenvalue}, we show that as $\mu\rightarrow \infty$, the eigenvalues of matrix $\langle F_N^T,\hat{H}F_N\rangle$ converges to the first $N$ eigenvalues of the Hamiltonian $\hat{H}$. This provides an affirmative answer to the conjecture stated in \cite{BIB:CMpaper1}.


First we prove the following lemmas. 
\begin{lemma}
Suppose that for $i,j=1,\ldots,N$
\begin{equation} \sum_{k=1}^{\infty} a_{ik}^*a_{jk}=\delta_{ij}. \label{equation:assumption}\end{equation}
Then, for any $k$,
\[  \sum_{i=1}^{N} |a_{ik}|^2\leq 1. \]
\label{lemma:aSquare}\end{lemma}
\proof It suffices to show the result for $k=1$ (i.e. by relabeling the indices, the result would follow for other $k$'s). Let $C$ be an $N\times N$ matrix whose $ij$-th entry is given by
\begin{equation} C_{ij}=\sum_{l=2}^{\infty}a^*_{il}a_{jl}. \label{equation:Cij}\end{equation}
By construction, $C$ is hermitian, we claim that it is also positive semi-definite matrix. For any vector $\x=(x_1,\ldots,x_N)^T$, observe that
\[ \x^* C \x=\sum_{i,j=1}^{N} x^*_iC_{ij}x_j=\sum_{i,j=1}^{N}x^*_ix_j\sum_{l=2}^{\infty}a^*_{il}a_{jl}=\sum_{l=2}^{\infty}\left ( \sum_{i=1}^N x_ia_{il}\right)^*   \left(  \sum_{j=1}^N x_ia_{jl}\right)= \sum_{l=2}^{\infty} \left| \sum_{i=1}^N x_ia_{il}\right|^2\geq 0.\] 
Using Cholesky decomposition, there exist a lower diagonal matrix $L$ (i.e. not necessarily unique as $C$ is semi-definite) such that $C=\bar{L}\bar{L}^\dagger$. Thus,
\[ C_{ij}=\sum_{m=1}^{N} L_{im}^*L_{jm}. \] 
In particular, comparing with equation \eqref{equation:Cij}, we conclude that for $i,j=1\ldots,N$,
 \[ \sum_{m=1}^{N} L_{im}^*L_{jm}=\sum_{l=2}^{\infty}a^*_{il}a_{jl} \]
For $i=1,\ldots,N$, set $L_{i0}:=a_{i1}$ and $\vec{L}_i=(L_{i0},L_{i1},\ldots,L_{iN})$ . From assumption \eqref{equation:assumption} and the above equality, we may conclude that 
\[ \langle \vec{L_i},\vec{L_j}\rangle=\delta_{ij}  \qquad \hbox{ for } \qquad i,j=1,\ldots,N. \]
Note that $\{\vec{L}_i\}_{i=1}^N$ are $(N+1)$-dimensional vectors. We can find vector $\vec{L_0}=(L_{00},L_{01},\ldots,L_{0N})$ such that $\{\vec{L}_i\}_{i=0}^N$ is an orthonormal set of basis. Form $(N+1)\times (N+1)$ matrix $M$ whose $ij$-th entry $M_{ij}$ is equal to $L_{ij}$ for $i,j=0,\ldots,N$. Observe that $M$ is hermitian as its rows are orthonormal. Hence columns of $M$ are orthonormal as well. In particular,
\[ 1=\sum_{i=0}^N |M_{i0}|^2=\sum_{i=0}^N |L_{i0}|^2=|L_{00}|^2+\sum_{i=1}^{N} |a_{i1}|^2 \Longrightarrow \sum_{i=1}^{N} |a_{i1}|^2=1-|L_{00}|^2\leq 1. \]
This completes the lemma. \qed

\begin{lemma}
Suppose $\lambda_N<\lambda_{N+1}$. For any $\epsilon>0$ there exist $\epsilon_0$ such that 
\[  \left|   \sum_{i=1}^{N}\langle g_i,\hat{H}g_i\rangle- \sum_{i=1}^{N}\langle \phi_i,\hat{H}\phi_i\rangle \right| <\epsilon_0 \qquad \hbox{ with }  \qquad \langle g_i,g_j\rangle=\delta_{i,j}\]
implies that $\|g_i-\varphi_i\|_2<\epsilon$, for $i=1\ldots,N$. Here $\varphi_1\ldots,\varphi_N$ is some unitary transformation of $\phi_1\ldots,\phi_N$.
\label{lemma:gphi}\end{lemma}

\proof Since $\{\phi_i\}_{i=1}^{\infty}$ form a complete set of basis in $L^2$, for every $i=1\ldots,N$, we have
\[  g_i=\sum_{k=1}^{\infty} a_{ik}\phi_k. \]
Moreover, the assumptions on the $g_i$'s imply that 
\begin{equation}  \sum_{k=1}^{\infty} a_{ik}^*a_{jk}=\delta_{ij}. \label{equation:aijL2}\end{equation}
By spectral decomposition,
\[  \hat{H}=\sum_{l=1}^{\infty} \lambda_l|\phi_l\rangle\langle\phi_l|.\]
Thus, for $i=1\ldots,N$
\[  \langle g_i,\hat{H} g_i\rangle=\langle g_i,\sum_{l=1}^{\infty} \lambda_l|\phi_l\rangle\langle\phi_l|  g_i\rangle=  \sum_{l=1}^{\infty}\lambda_l \langle g_i,\phi_l \rangle  \langle \phi_l,g_i \rangle=  \sum_{l=1}^{\infty}\lambda_l|a_{il}|^2.\]
Summing over $i$ on both sides of the above equality, we have 
\[ \sum_{i=1}^{N}\langle g_i,\hat{H} g_i\rangle=\sum_{i=1}^{N}   \sum_{l=1}^{\infty}\lambda_l|a_{il}|^2=\sum_{l=1}^{\infty}\lambda_l\sum_{i=1}^{N}|a_{il}|^2=\sum_{l=1}^{\infty} \lambda_l b_l, \]
where
\[ b_l:=\sum_{i=1}^{N}|a_{il}|^2. \]
Observe that $b_l$'s satisfy the following properties:
\begin{equation}
0\leq b_l \leq 1 \quad \hbox{ for } \quad l=1,\ldots, 
\label{equation:bProperty1}\end{equation} 
and
\begin{equation}  \sum_{l=1}^\infty b_l=N.
\label{equation:bProperty2} \end{equation}
We can see the first property using lemma \ref{lemma:aSquare}. To see the second property note that 
\[ \sum_{l=1}^\infty b_l=\sum_{l=1}^\infty \sum_{i=1}^{N}|a_{il}|^2=  \sum_{i=1}^{N}\sum_{l=1}^\infty |a_{il}|^2= \sum_{i=1}^{N} 1=N. \]
Now observe that 
\begin{align}
 &\sum_{i=1}^{N}\langle g_i,\hat{H}g_i\rangle- \sum_{i=1}^{N}\langle \phi_i,\hat{H}\phi_i\rangle \notag \\
=& \sum_{l=1}^{\infty}\lambda_l b_l-(\lambda_1+\cdots+\lambda_N)  \notag \\
\geq & (b_1-1)\lambda_1+\cdots+(b_N-1)\lambda_N+(\sum_{l=N+1}^{\infty} b_l) \lambda_{N+1}  \notag \\
= &   (b_1-1)\lambda_1+\cdots+(b_N-1)\lambda_N+(N-b_1-\cdots-b_N) \lambda_{N+1}  \notag \\
=& (1-b_1)(\lambda_{N+1}-\lambda_1)+\cdots+ (1-b_N)(\lambda_{N+1}-\lambda_N), \label{equation:summation}
\end{align}
where we used the nondecreasing ordering of $\lambda_i$'s in the third line, and \eqref{equation:bProperty2} in the fourth line. Now from \eqref{equation:bProperty1} and  nondecreasing ordering of $\lambda_i$'s, each of the terms in summation \eqref{equation:summation} is positive. Hence 
\[  \left |\sum_{i=1}^{N}\langle g_i,\hat{H}g_i\rangle- \sum_{i=1}^{N}\langle \phi_i,\hat{H}\phi_i\rangle \right| \geq (1-b_1)(\lambda_{N+1}-\lambda_1)+\cdots+ (1-b_N)(\lambda_{N+1}-\lambda_N). \]
Now using the assumption that $\lambda_{N+1}$ is strictly greater than $\lambda_1,\ldots,\lambda_N$, we may conclude that for every $\epsilon_1>0$, there exist $\epsilon_0$ such that if the LHS of the above inequality is smaller than $\epsilon_0$, then 
\[  1-b_l < \epsilon_1 \quad \hbox{ for } \quad l=1,\ldots,N. \]
Moreover, using equation \eqref{equation:bProperty2}, we can conclude that 

\[ \sum_{l=N+1}^\infty b_l < N\epsilon_1 \Longrightarrow \sum_{l=N+1}^{\infty}\sum_{i=1}^{N} |a_{il}|^2=\sum_{i=1}^{N}  \sum_{l=N+1}^{\infty} |a_{il}|^2<N\epsilon_1.\]
In particular for $i=1\ldots,N$, we have 
\begin{equation}
\sum_{l=N+1}^{\infty} |a_{il}|^2<N\epsilon_1. 
\label{equation:Nepsilon1}\end{equation}

Next we show that $N\times N$ matrix $\{a_{ik}\}_{i,k=1}^N$ is ``almost'' unitary in the sense that 

\begin{align} \left|\delta_{ij}- \sum_{k=1}^{N} a_{ik}^*a_{jk}  \right|=\left| \sum_{k=N+1}^{\infty} a_{ik}^*a_{jk}\right|& \leq \left(  \sum_{k=N+1}^{\infty} |a_{ik}|^2 \right)^{1/2}\left(  \sum_{k=N+1}^{\infty} |a_{jk}|^2 \right)^{1/2}   \notag \\
& < (N\epsilon_1)^{1/2}(N\epsilon_1)^{1/2} =N\epsilon_1, \notag
\end{align}

where we used \eqref{equation:aijL2} for the first equality, Cauchy-Schwarz for the first inequality, and \eqref{equation:Nepsilon1} for the second inequality. We can orthonormalize the rows of matrix  $\{a_{ik}\}_{i,k=1}^N$ using Gram-Schmidt process, to form a new unitary matrix $\{a'_{ik}\}_{i,k=1}^N$ . Indeed, because of the above inequality, for any $\epsilon_2>0$, we can choose $\epsilon_1$ small enough such that 
\begin{equation} \sum_{k=1}^{N} |a_{ik}-a'_{ik}|^2<\epsilon_2 \qquad \hbox{ for every }\qquad i=1,\ldots,N.\label{equation:epsilon2}\end{equation} 
Now, for $i=1,\dots,N$, set
\[  \varphi_i=\sum_{k=1}^{N} a'_{ik}\phi_k. \] 
Observe that 
\[  \|g_i-\varphi_i\|_2^2=\sum_{k=1}^{N} |a_{ik}-a'_{ik}|^2+\sum_{k=N+1}^{\infty}|a_{ik}|^2 <\epsilon_2+N\epsilon_1,\]
where we used \eqref{equation:epsilon2} and \eqref{equation:Nepsilon1} for the inequality.

Hence, for any $\epsilon>0$, we can choose $\epsilon_1$ and $\epsilon_2$ small enough such that  $\|g_i-\varphi_i\|_2<\epsilon$ for $i=1,\ldots,N$. The result follows. \qed

Now, we prove the main result of this section.
\begin{theorem}
Assume $\lambda_{N+1}>\lambda_N$. For every $\epsilon>0$, there exist $\mu_0$ such that for $\mu>\mu_0$, the solutions to the regularized optimization problem \eqref{equation:propertyF} satisfy
\[ \|f_i-\varphi_i\|_2<\epsilon \qquad \hbox{ for } \qquad i=1\ldots,N, \]
where $\varphi_1\ldots,\varphi_N$ is some unitary transformation of $\phi_1\ldots,\phi_N$.
\label{theorem:main}\end{theorem}
For given $\epsilon$, choose $\epsilon_0$ as indicated by lemma \ref{lemma:gphi}. Choose $\mu_0$ large enough such that 
\begin{equation} \frac{1}{\mu_0}\sum_{i=1}^{N}J(\phi_i) <\epsilon_0. 
\label{equation:mu0thm1}\end{equation}
Let $\mu>\mu_0$. Observe that 
\[
 \sum_{i=1}^{N}\langle \phi_i,\hat{H}\phi_i\rangle 
\leq   \sum_{i=1}^{N}\langle f_i,\hat{H}f_i\rangle
\leq   \frac{1}{\mu}\sum_{i=1}^{N}J(f_i)+ \sum_{i=1}^{N}\langle f_i,\hat{H}f_i\rangle 
\leq   \frac{1}{\mu}\sum_{i=1}^{N}J(\phi_i)+ \sum_{i=1}^{N}\langle \phi_i,\hat{H}\phi_i\rangle.
\]
where we used property \eqref{equation:propertyPhi} for the first inequality, and property \eqref{equation:propertyF} for the last inequality. Hence, using \eqref{equation:mu0thm1}, we have
\[
 \sum_{i=1}^{N}\langle \phi_i,\hat{H}\phi_i\rangle 
\leq   \sum_{i=1}^{N}\langle f_i,\hat{H}f_i\rangle
\leq  \epsilon_0+ \sum_{i=1}^{N}\langle \phi_i,\hat{H}\phi_i\rangle,
\]
which implies that 
\begin{equation}  \left|\sum_{i=1}^{N}\langle f_i,\hat{H}f_i\rangle- \sum_{i=1}^{N}\langle \phi_i,\hat{H}\phi_i\rangle \right| <\epsilon_0. \label{equation:sumFConvergeSumPhi} \end{equation}
Now applying lemma \ref{lemma:gphi}, completes the proof.\qed

\begin{remark}
The assumption $\lambda_{N+1}>\lambda_N$ is essential. Because otherwise, we can have situations in which $\psi_N$ converges to $\phi_{N+1}$ in $L^2$ norm and the result of theorem \ref{theorem:main} would clearly not hold. 
\label{remark:essentialAssumption}\end{remark}

\begin{theorem}
Eigenvalues $(\nu_1,\ldots,\nu_N)$ of matrix $\langle F^T,\hat{H} F\rangle$, converge to $(\lambda_1,\ldots,\lambda_N)$, the eigenvalues of Hamiltonian $\hat{H}$.
\label{theorem:mainEigenvalue}\end{theorem}
\proof Using the same notation as before, let $\varphi_i$, for $i=1,\ldots,N$, be the unitary transformation of the eigenfunctions $\phi_1,\ldots,\phi_N$ as described by theorem \ref{theorem:main}. Let $\langle\Phi^T,\hat{H}\Phi\rangle$ be an $N\times N$ matrix with $(j,k)$-th entry equal to $\langle\varphi_j,\hat{H}\varphi_k\rangle$. We show that matrix $\langle F^T,\hat{H}F\rangle$ converges to matrix $\langle\Phi^T,\hat{H}\Phi\rangle$ in Frobenius norm. For matrix $\langle\Phi^T,\hat{H}\Phi\rangle$ has the same eigenvalues $(\lambda_1,\ldots,\lambda_N)$ as the Hamiltonian, since $\varphi_i$'s are unitary transformation of the eigenfunctions $\phi_i$'s. To that end, it suffices to show that
\[ \langle f_i,\hat{H} f_j\rangle\rightarrow \langle \varphi_i,\hat{H}\varphi_j\rangle \qquad \hbox{ for } i,j=1,\ldots,N. \] 

Suppose 
\[ \varphi_i=\sum_{k=1}^{N} a'_{ik}\phi_k  \qquad \hbox{ for } i=1,\ldots,N,\]
and 
\[ f_i=\sum_{k=1}^{\infty} a_{ik}\phi_k  \qquad \hbox{ for } i=1,\ldots,N.\]
From the result of theorem \ref{theorem:main} we know that 
\begin{equation} \|f_i-\varphi_i\|^2_2=\sum_{k=1}^{N} |a_{ik}-a'_{ik}|^2+\sum_{k=N+1}^{\infty}|a_{ik}|^2 \rightarrow 0, \qquad \hbox{ as } \qquad \mu\rightarrow \infty.\label{equation:fphi0}\end{equation}
This also implies that as $\mu\rightarrow \infty$, for $i=1\ldots,N$, 
\begin{equation} a_{ik}\rightarrow a'_{ik}, \quad \hbox{ where } \quad 1\leq k\leq N,\qquad \hbox{ and } \qquad a_{ik}\rightarrow 0 \quad \hbox{ where } \quad k>N.
\label{equation:aConverge}\end{equation}
Since $\sum_{i=1}^N \langle \phi_i,\hat{H} \phi_i \rangle=\sum_{i=1}^{N}\langle \varphi_i,\hat{H}\varphi_i\rangle$, from the proof of theorem \ref{theorem:main} (i.e. equation \eqref{equation:sumFConvergeSumPhi}), we can conclude that 
\[ \sum_{i=1}^{N} \langle f_i,\hat{H} f_i\rangle \rightarrow \sum_{i=1}^{N} \langle \varphi_i,\hat{H}\varphi_i \rangle \qquad \hbox{ as }  \qquad \mu\rightarrow \infty.\]
Rewriting the above expression we have 
\[ \sum_{i=1}^{N} \sum_{k=1}^{\infty} \lambda_k |a_{ik}|^2 \rightarrow \sum_{i=1}^{N} \sum_{k=1}^{N} \lambda_k |a'_{ik}|^2 \qquad \hbox{ as } \qquad \mu\rightarrow \infty.\]
From \eqref{equation:aConverge} we can conclude that for any finite $M>N$, $\sum_{i=1}^{N} \sum_{k=1}^{M} \lambda_k |a_{ik}|^2$ converges to $\sum_{i=1}^{N} \sum_{k=1}^{N} \lambda_k |a'_{ik}|^2$ as $\mu\rightarrow\infty$.
Hence, 
\begin{equation} \sum_{i=1}^{N} \sum_{k=M}^{\infty} \lambda_k |a_{ik}|^2 \rightarrow 0 \qquad \mu\rightarrow \infty. \label{equation:a0}\end{equation}
Now for $i,j=1,\ldots,N$, consider 
\begin{align}
 \langle f_i,\hat{H} f_j\rangle-\langle \varphi_i,\hat{H} \varphi_j\rangle &=\sum_{k=1}^{\infty}\lambda_ka^*_{ik}a_{jk}- \sum_{k=1}^{N}\lambda_k a'^*_{ik}a'_{jk} \notag \\
& = \sum_{k=1}^{N}\lambda_k (a^*_{ik}a_{jk}-a'^*_{ik}a'_{jk})+\sum_{k=N+1}^{\infty}\lambda_ka^*_{ik}a_{jk} \label{equation:line2}
\end{align} 
The first summation in line \eqref{equation:line2} goes to zero as $\mu\rightarrow \infty$, because of \eqref{equation:aConverge}. If all $\lambda_k$'s are negative (i.e. and therefore bounded, as we have arranged them in nondecreasing order), then a simple application of Cauchy-Schwarz and using \eqref{equation:fphi0}, implies that the second summation in line \eqref{equation:line2} also goes to zeros as $\mu\rightarrow \infty$. Otherwise, there exist finite $M$ greater than $N$ such that for $k>M$, $\lambda_k$ is positive. Now we write the second summation in line \eqref{equation:line2} as 
\[ \sum_{k=N+1}^{M}\lambda_ka^*_{ik}a_{jk}+\sum_{k=M+1}^{\infty}\lambda_ka^*_{ik}a_{jk}.\]
Again the first summation in above line goes to zero as $\mu\rightarrow \infty$ because of \eqref{equation:aConverge}. For the second summation in the above expression, note that by Cauchy-Schwarz
\[  \left|\sum_{k=M+1}^{\infty}\lambda_ka^*_{ik}a_{jk} \right| \leq \left( \sum_{k=M+1}^{\infty}\lambda_k|a_{ik}|^2 \right)^{1/2} \left( \sum_{k=M+1}^{\infty}\lambda_k|a_{jk}|^2 \right)^{1/2}. \]
The reason that we use $\lambda_k$ instead of $|\lambda_k|$ on the RHS is due to the assumption that $\lambda_k$'s are positive for $k>M$. Equation \eqref{equation:a0}, in particular, yields that the RHS of the above expression goes to zero. Thus, we have shown that the expression on line \eqref{equation:line2} goes to 0 as $\mu\rightarrow \infty$. This completes the proof. \qed

\begin{remark}
Observe that theorem \ref{theorem:mainEigenvalue} does not immediately follow from the result of theorem \ref{theorem:main}. This is due to the fact that the Hamiltonian $\hat{H}$ is not generally a bounded operator on $L^2$ functions. 
\label{remark:difficult}\end{remark}

\section{Conclusion}\label{section:conclusion}
In this paper we prove consistency results for compressed modes introduced in \cite{BIB:CMpaper1}. Although we show the results hold in a more general setting, the most important application of the results of this paper is for compressed modes.

In \cite{BIB:CMpaper1,BIB:CMpaper2}, the authors pioneered the use of $L^1$ regularization to compute modes that are spatially localized. We proved that, under some necessary assumptions on the spectrum of the Hamiltonian, as the regularization term in the non-convex optimization problem \eqref{equation:propertyPsi} vanishes, the compressed modes indeed converge to Wannier-like functions. We also provided an affirmative proof for a conjecture in \cite{BIB:CMpaper1}.

\section{Acknowledgment}\label{section:acknowledgment}
I am indebted to Professors Russel Caflisch and Stan Osher for invaluable guidance and helpful discussions and comments. The research was partially supported by DOE grant number DE-SC0010613, and NSERC PGS-D award.

\end{document}